# Caveats for the use of Web of Science Core Collection in old literature retrieval and historical bibliometric analysis

Weishu Liu    wsliu08@163.com   https://orcid.org/0000-0001-8780-6709

School of Information Management and Artificial Intelligence, Zhejiang University of Finance and Economics, Hangzhou 310018, Zhejiang, China

**Abstract** By using publications from Web of Science Core Collection (WoSCC), Fosso Wamba and his colleagues published an interesting and comprehensive paper in Technological Forecasting and Social Change to explore the structure and dynamics of artificial intelligence (AI) scholarship. Data demonstrated in Fosso Wamba's study implied that the year 1991 seemed to be a "watershed" of AI research. This research note tried to uncover the 1991 phenomenon from the perspective of database limitation by probing the limitations of search in abstract/author keywords/keywords plus fields of WoSCC empirically. The low availability rates of abstract/author keywords/keywords plus information in WoSCC found in this study can explain the "watershed" phenomenon of AI scholarship in 1991 to a large extent. Some other caveats for the use of WoSCC in old literature retrieval and historical bibliometric analysis were also mentioned in the discussion section. This research note complements Fosso Wamba and his colleagues' study and also helps avoid improper interpretation in the use of WoSCC in old literature retrieval and historical bibliometric analysis.

## 1.  Introduction

Recently, Fosso Wamba and his colleagues published an interesting and comprehensive paper in Technological Forecasting and Social Change to explore the structure and dynamics of artificial intelligence (AI) scholarship and investigate whether we were preparing for a good AI society (Fosso Wamba et al. 2021). The above study used "artificial intelligence" as the search term to search in the topic field (a combination of title, abstract, author keywords and keywords plus fields) of Web of Science Core Collection (WoSCC). The variable publication year was used widely throughout this paper. The data demonstrated in Fosso Wamba et al. (2021) implied that the year 1991 seemed to be a "watershed" of AI research. For example, according to Fosso Wamba et al. (2021), AI publications at least doubled in 1991; most of the top fifty influential AI papers were published after 1991; and AI for social good publications started from 1991. However, in order to be more precise and uncontroversial, some possible explanations are needed to interpret the abnormal 1991 phenomenon in AI scholarship.

By using the newly available search field tags provided by WoSCC, this research note explored the limitations of search in abstract/author keywords/keywords plus fields empirically[1]. The low availability rates of abstract/author keywords/keywords plus information in WoSCC found in this study can explain the "watershed" phenomenon of AI scholarship in 1991 to a large extent. This empirical research will make Fosso Wamba and his colleagues' paper more comprehensive and understandable. More general users of WoSCC, especially those outside the field of Scientometrics, will have a more comprehensive understanding of the limitations of this widely used bibliographic database (González-Alcaide 2021; Li et al. 2018; Zhu & Liu 2020). At the same time, this study can also help avoid improper interpretation in the use of WoSCC in old literature retrieval and historical bibliometric analysis.

## 2. Data and methods

This study explored the availability of abstract/author keywords/keywords plus information in three classical journal citation indexes of WoSCC (i.e., Science Citation Index Expanded, Social Sciences Citation Index, and Arts & Humanities Citation Index). Three newly available field tags (AB for the abstract field, AK for the author keywords field, and KP for the keywords plus field[2]) enabled the online large-scale information availability checking. The following three search queries were used to test if a record had abstract/author keywords/keywords plus information in WoSCC. Same with Fosso Wamba et al. (2021), we also first chose the data of the latest 45 years for investigation by setting the timespan as 1975 to 2019 in advanced search platform of WoSCC[3]. Data were accessed on 5th February 2021 via the library of Sichuan University (Sichuan University has subscribed Science Citation Index Expanded from 1900, Social Sciences Citation Index from 1900, and Arts & Humanities Citation Index from 1975).

> AB=(A* OR B* OR C* OR D* OR E* OR F* OR G* OR H* OR I* OR J* OR K* OR L* OR M* OR N* OR O* OR P* OR Q* OR R* OR S* OR T* OR U* OR V* OR W* OR X* OR Y* OR Z* OR 0* OR 1* OR 2* OR 3* OR 4* OR 5* OR 6* OR 7* OR 8* OR 9*)
>
> AK=(A* OR B* OR C* OR D* OR E* OR F* OR G* OR H* OR I* OR J* OR K* OR L* OR M* OR N* OR O* OR P* OR Q* OR R* OR S* OR T* OR U* OR V* OR W* OR X* OR Y* OR Z* OR 0* OR 1* OR 2* OR 3* OR 4* OR 5* OR 6* OR 7* OR 8* OR 9*)
>
> KP=(A* OR B* OR C* OR D* OR E* OR F* OR G* OR H* OR I* OR J* OR K* OR L* OR M* OR N* OR O* OR P* OR Q* OR R* OR S* OR T* OR U* OR V* OR W*

---

[1] We didn't test the availability rate of records' title information in WoSCC for three reasons. First of all, most of the records were with title information in WoSCC according to the author's observation. Secondly, for some records without title information, WoSCC has filled the title field in the database with the word "untitled". And lastly, WoSCC didn't support the use of the same search queries in Data and methods section in the title field.

[2] According to Clarivate, keywords plus "are index terms automatically generated from the titles of cited articles" by WoSCC. Accessed on 4th March 2021 via https://images.webofknowledge.com/images/help/WOS/hp_full_record.html. For more information about keywords plus, please refer to https://support.clarivate.com/ScientificandAcademicResearch/s/article/KeyWords-Plus-generation-creation-and-changes?language=en_US.

[3] http://apps.webofknowledge.com/WOS_AdvancedSearch_input.do?product=WOS&SID=8E9ASl1Riz7rxClXxzW&search_mode=AdvancedSearch

OR X* OR Y* OR Z* OR 0* OR 1* OR 2* OR 3* OR 4* OR 5* OR 6* OR 7* OR 8* OR 9*)

## 3. Availability of abstract/author keywords/keywords plus information in WoSCC

3.1. All document types considered

We first took the Science Citation Index Expanded (SCIE) as a case to explore the availability of abstract/author keywords/keywords plus information in WoSCC. About 9.15 million records were indexed by SCIE during the period 1975-1989, however, only about eight thousand of them (0.09%) were available with abstract information in WoSCC. The number of records with abstract information in WoSCC rose from 3375 in 1989 to 52973 in 1990 and further soared to 469325 in 1991. Correspondingly, the share of records with abstract information in WoSCC rose from 0.5% in 1989 to 7.7% in 1990 and further exploded to 66.0% in 1991. Since then, the share maintained a steady upward trend until it reached a new high point in 2019 (77.3%).

Similar trends reoccurred in the author keywords and keywords plus fields. Very limited shares of SCIE indexed records published before 1990 were with author keywords/keywords plus information in WoSCC. As demonstrated in Figure 1, the share of records with author keywords information in WoSCC was much lower than the share of records with abstract information for the period 1991-2019. Only 2.9% of SCIE indexed records published in 1990 were with author keywords information in WoSCC and only 24.8% of them were with author keywords information in WoSCC for the year 1991. After then, the share rose linearly from 24.8% in 1991 to 63.2% in 2019. Comparatively, the share of records with keywords plus information in WoSCC was only a bit lower than the share of records with abstract information for the period 1991-2019.

For Social Sciences Citation Index (SSCI) indexed records, similar results can be got. However, low availability rates of abstract/author keywords/keywords plus information for Arts & Humanities Citation Index (A&HCI) indexed records can be found, especially for records published before 2000. After then, the availability rates of the three fields for A&HCI indexed records rose gradually but were still much lower than those of SCIE and SSCI indexed records. For brevity, the details of the availability of abstract/author keywords/keywords plus information in WoSCC for SSCI and A&HCI indexed records were not demonstrated in this article.

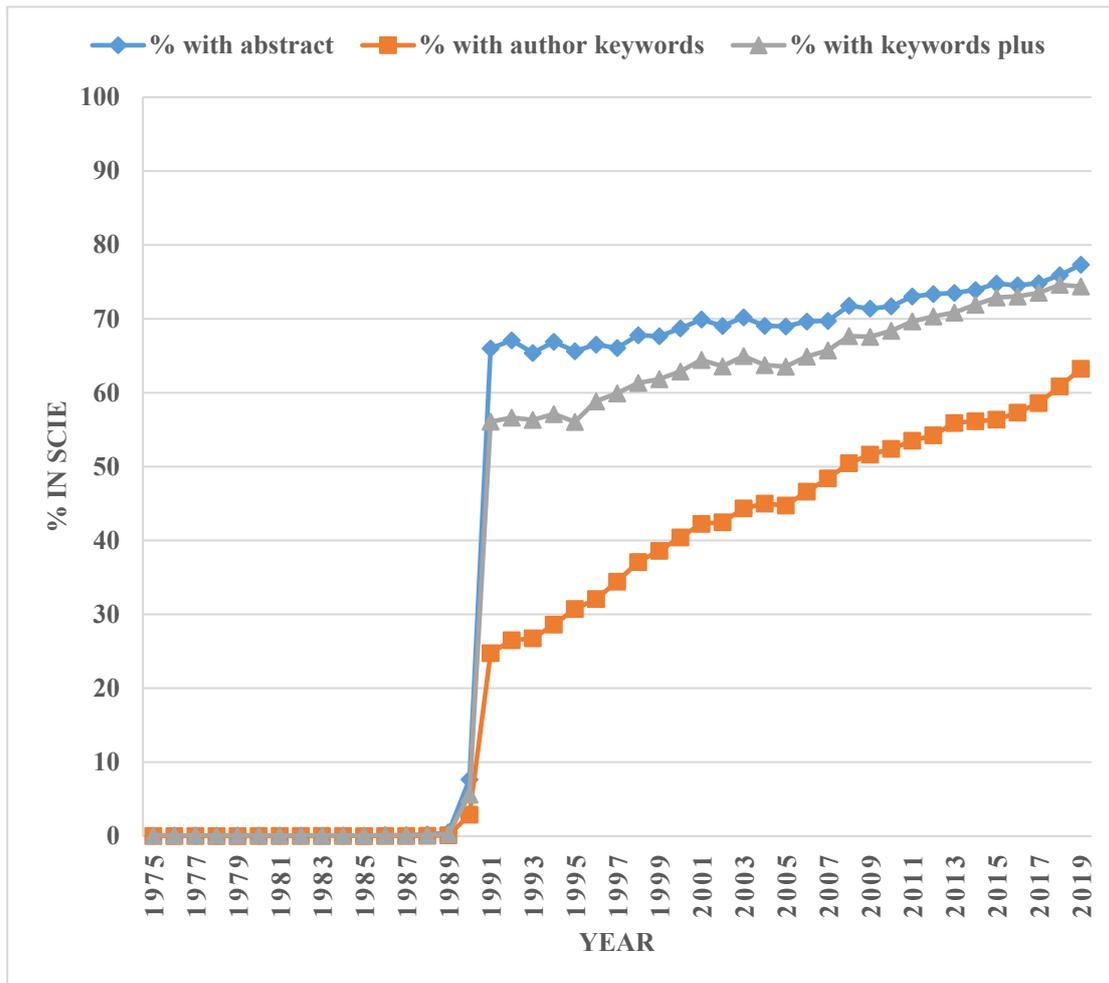

Figure 1 Availability of abstract/author keywords/keywords plus information in WoSCC for all SCIE indexed records

3.2. Only articles and reviews considered

3.2.1 Science Citation Index Expanded scenario

Articles and reviews were two most widely used document types in literature retrieval and bibliometric analysis. Figure 2 also demonstrated the availability of abstract/author keywords/keywords plus information in WoSCC for SCIE indexed articles and reviews. The dynamics of the availability of the three fields for articles and reviews were similar to that of all SCIE indexed records as demonstrated in Figure 1. For all the three fields, less than 1% of SCIE indexed articles and reviews published before 1990 were with corresponding information available in WoSCC. Significant growth happened for the year 1990. 9.7%, 3.7%, and 7.0% of these records published in 1990 were with abstract/author keywords/keywords plus information in WoSCC respectively. The share of articles and reviews with abstract information in WoSCC rose from 82.1% in 1991 to 99.1% in 2019. Comparatively, the share of articles and reviews with author keywords information was much lower. Only 30.5% of articles and reviews were with author keywords information available in WoSCC for the year 1991. Although the share rose rapidly since then, less than half of SCIE indexed articles and reviews published before 1999 were with author keywords information in WoSCC and the share accounted for 79.7% for the year 2019. Comparatively, the share of articles and reviews with keywords plus information in WoSCC rose from 66.8% in 1991 to 94.3% in 2018 but dropped slightly to 93.1% in 2019.

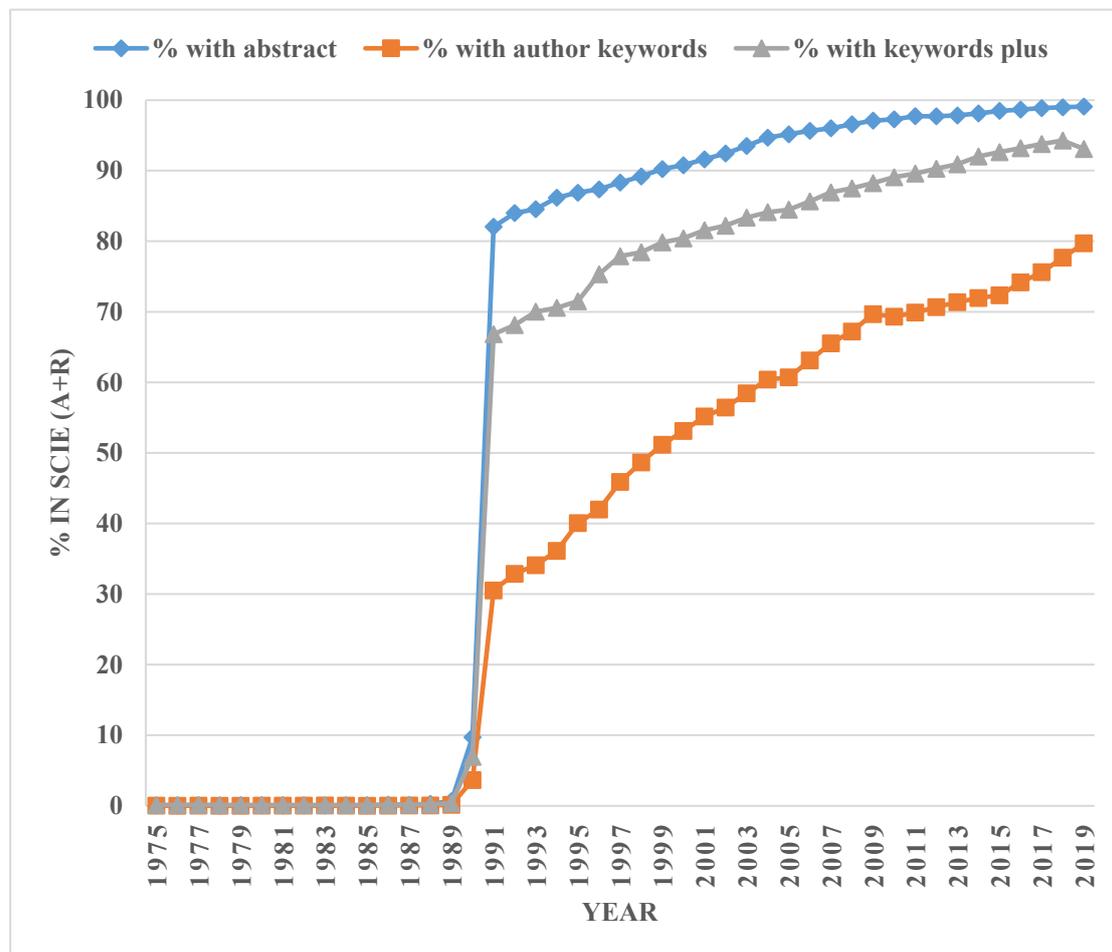

Figure 2 Availability of abstract/author keywords/keywords plus information in WoSCC for SCIE indexed articles and reviews
Note: (A+R) refers to (Articles + Reviews)

3.2.2 Social Sciences Citation Index scenario

Figure 3 demonstrated the availability of abstract/author keywords/keywords plus information in WoSCC for SSCI indexed articles and reviews. For all the three fields, less than 0.5% of SSCI indexed articles and reviews published before 1990 were with corresponding information available in WoSCC. The share of articles and reviews with abstract information rose to 2.4% in 1990, rocketed to 29.1% in 1991 and further doubled to 65.4% in 1992. After then, the share rose gradually to 97.9% in 2019. Comparatively, the share of articles and reviews with author keywords information in WoSCC rose significantly from 2.0% in 1990 to 14.9% in 1991 and since then rose gradually to 84.0% in 2019. Similar to the SCIE indexed articles and reviews, the share of SSCI indexed articles and reviews with author keywords information was also lower compared to the share with abstract information for the period 1991-2019. Less than half of SSCI indexed articles and reviews were with author keywords information in WoSCC for records published before 2005. The share of articles and reviews with keywords plus information in WoSCC rose significantly from 5.9% in 1990 to 49.3% in 1991. After then, the share rose gradually to 91.3% in 2019.

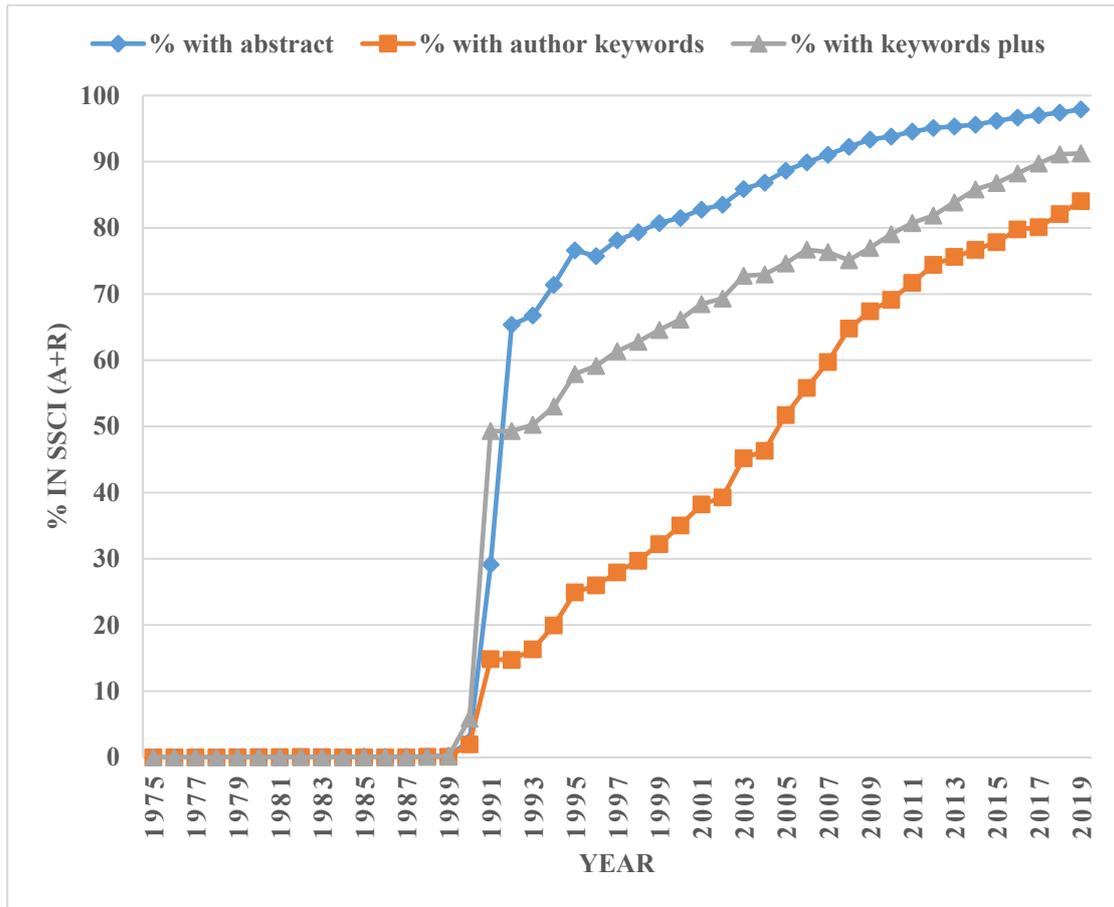

Figure 3 Availability of abstract/author keywords/keywords plus information in WoSCC for SSCI indexed articles and reviews

Note: (A+R) refers to (Articles + Reviews)

3.2.3 Arts & Humanities Citation Index scenario

For all the three fields, less than 0.5% of A&HCI indexed articles and reviews published before 1990 were with corresponding information available in WoSCC. The share of articles and reviews with abstract information continued to be very low in 1990 (0.1%) but rose to 2.3% in 1991 and further rose to 8.9% in 1992. However, the share dropped slightly during the next six years but rerose significantly from 5.6% in 1998 to 73.6% in 2019. Similarly, the shares of articles and reviews with author keywords and keywords plus information went through the period of growth, flatness and regrowth as demonstrated in Figure 4. Compared with SCIE and SSCI indexed articles and reviews, the shares of A&HCI indexed articles and reviews with corresponding information were much lower for the period 1991-2019. Even for the year 2019, only 55.9% and 33.6% of A&HCI indexed articles and reviews were with author keywords and keywords plus information in WoSCC respectively.

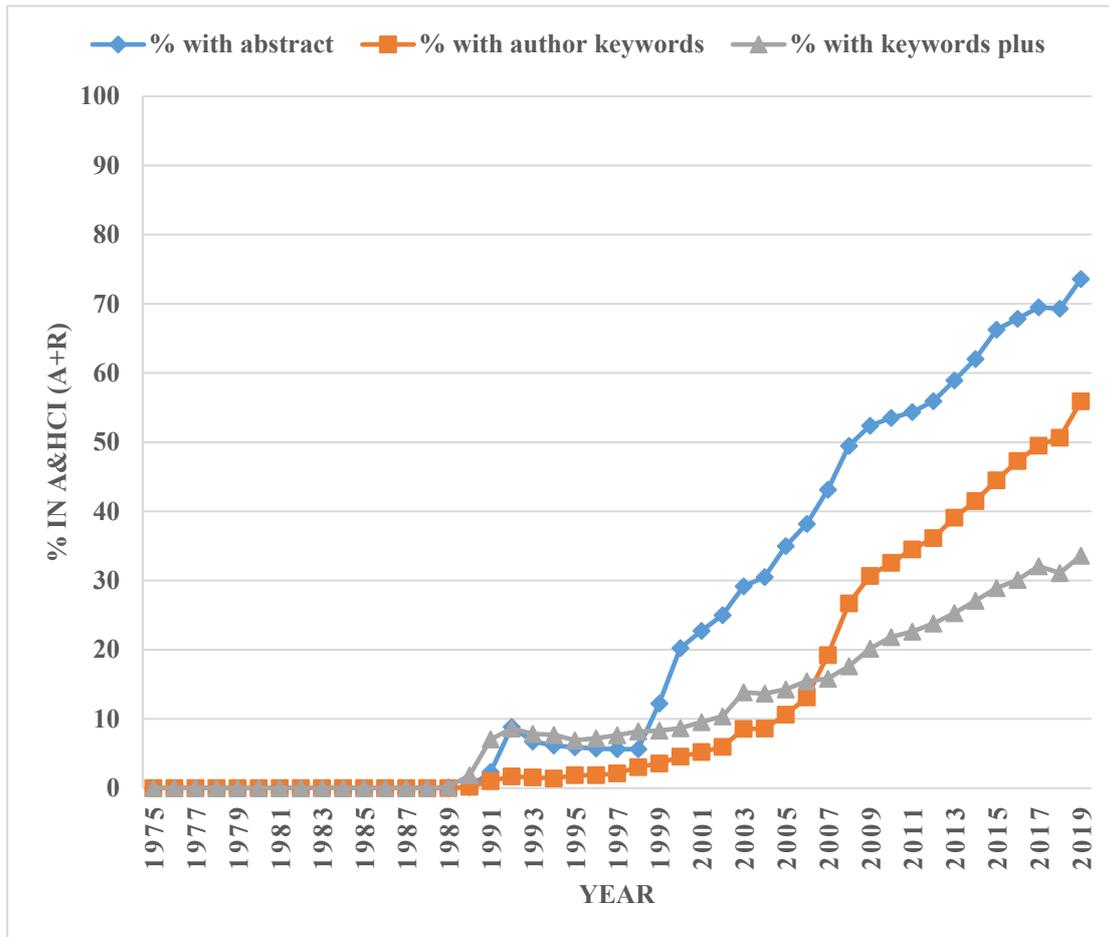

Figure 4 Availability of abstract/author keywords/keywords plus information in WoSCC for A&HCI indexed articles and reviews

Note: (A+R) refers to (Articles + Reviews)

## 4. Overview of the situation prior to 1975

Both SCIE and SSCI cover records published from 1900 to present, however, A&HCI covers records published from 1975 to present (Birkle et al. 2020). As suggested by one of the reviewers, we further checked the availability of abstract/author keywords/keywords plus information in WoSCC for SCIE and SSCI indexed records published prior to 1975. Data in this section were accessed on July 6[th] 2021 via the library of Sichuan University.

4.1. Science Citation Index Expanded scenario

For 6.32 million SCIE indexed records published during 1900 and 1974, only 4810, 465, and 4112 of them were with abstract/author keywords/keywords plus information collected in WoSCC respectively. Even for 4.16 million SCIE indexed articles and reviews, only 0.11%, 0.01%, and 0.10% of them were with abstract/author keywords/keywords plus information collected in WoSCC respectively. Besides, based on the data we found that the availability rates of abstract/author keywords/keywords plus information in WoSCC remained extremely low throughout the period from 1900 to 1974.

### 4.2. Social Sciences Citation Index scenario

The low availability rates of abstract/author keywords/keywords plus information of SSCI indexed records in WoSCC were also evident for the whole studied period from 1900 to 1974. For 1.77 million SSCI indexed records, only 56 records were with abstract information in WoSCC, 1589 records were with keywords plus information in WoSCC, and none of all the 1.77 million records were with author keywords information collected in WoSCC. Even for 0.87 million SSCI indexed articles and reviews, less than 0.01% of them were with abstract information and 0.18% of them were with keywords plus information in WoSCC.

## 5. Discussion

The newly available topic related search field tags provided by WoSCC enabled the large-scale empirical exploration of the availability of abstract/author keywords/keywords plus information in WoSCC and further gave a reasonable explanation for the "watershed" phenomenon in 1991 for AI research which was implied by the data in Fosso Wamba's study (Fosso Wamba et al. 2021). This large-scale empirical examination promoted previous meaningful explorations (Fu et al. 2010; Heneberg 2011a, 2011b, 2013; Ho 2013; Pautasso 2014) and also provided more accurate and comprehensive information about the limitations of the abstract/author keywords/keywords plus fields in WoSCC than that provided in Clarivate's official help file (Clarivate 2020).

The extremely low availability rates of abstract/author keywords/keywords plus information before 1990 were evidenced empirically for SCIE, SSCI, and A&HCI indexed records. Even for articles and reviews, the availability rate of author keywords field was relatively low for records published during 1990s in all the three citation indexes. The A&HCI indexed records still suffered from the relatively lower availability rates of abstract/author keywords/keywords plus information even for most recent publications. Luckily, the low availability rates of abstract/author keywords/keywords plus information in WoSCC has gradually improved over the past two decades.

The low availability rates of abstract/author keywords/keywords plus information for early publications in WoSCC can be explained by the following causes. For early periods especially before 1990, WoSCC has not collected corresponding information systematically from journal publishers. WoSCC may also omit corresponding information to some extent after then. On the other hand, some publications themselves, especially for many non-substantial items, may have no abstract/author keywords information. Similarly, WoSCC may not have enough reference data to generate corresponding keywords plus information.

The low availability rates of abstract/author keywords/keywords plus information for early publications in WoSCC and even most recent records in A&HCI will make them less visible in literature retrieval and analysis. Many old and classical literature may be omitted in initial literature retrieval process and further be underestimated in analysis process simply due to the omission of abstract/author keywords/keywords plus information in WoSCC. Without proper treatment, the reliability of topic/keyword analysis based on these fields will also be hindered.

Apart from the abstract/author keywords/keywords plus fields, some other features of WoSCC will also influence old literature retrieval and historical bibliometric analysis. If we don't fully grasp the characteristics of author's affiliation (Jacsó 2009; Liu et al. 2018) and funding acknowledgment information (Liu et al. 2020; Paul-Hus et al. 2016; Tang et al. 2017) in WoSCC, corresponding literature retrieval and analysis may be biased. Besides, full author names were captured in WoSCC since June 2006 (Clarivate 2018), serious name ambiguity problem exists for records published before 2006, especially for authors from East Asia (Harzing 2015; Tang & Walsh 2010). Furthermore, the changing coverage of regional journals especially non-English journals in WoSCC may also influence the interpretation of related studies (Liu 2017; Vera-Baceta et al. 2019). And finally, since different institutions may subscribe to customized version of WoSCC, it is also important to detail the sub-datasets and corresponding coverage timespans of the used sub-datasets, especially for old literature retrieval and historical bibliometric analysis (Calver et al. 2017; Liu 2019).

As suggested by one of the reviewers, studies are expected to develop strategies that could alleviate these issues, especially by using some cutting-edge technologies. For Scientometrics researchers, what we can do now is broadcasting the main findings of this study to more and more users of WoSCC and readers of WoSCC-based studies.

**Conflicts of interest**: The author declares no conflict of interest.

**References**

Birkle, C., Pendlebury, D. A., Schnell, J., & Adams, J. (2020). Web of Science as a data source for research on scientific and scholarly activity. *Quantitative Science Studies, 1*(1), 363–376. https://doi.org/10.1162/qss_a_00018

Calver, M. C., Goldman, B., Hutchings, P. A., & Kingsford, R. T. (2017). Why discrepancies in searching the conservation biology literature matter. *Biological Conservation*, *213*, 19-26. https://doi.org/10.1016/j.biocon.2017.06.028

Clarivate (2018). https://support.clarivate.com/ScientificandAcademicResearch/s/article/Web-of-Science-Core-Collection-Explanation-on-Full-Author-Names?language=en_US accessed on 3rd March 2021.

Clarivate (2020). https://images.webofknowledge.com/images/help/WOS/hp_full_record.html accessed on 3rd March 2021.

Fosso Wamba, S. F., Bawack, R. E., Guthrie, C., Queiroz, M. M., & Carillo, K. D. A. (2021). Are we preparing for a good AI society? A bibliometric review and research agenda. *Technological Forecasting and Social Change*, 164, 120482. https://doi.org/10.1016/j.techfore.2020.120482

Fu, H. Z., Ho, Y. S., Sui, Y. M., & Li, Z. S. (2010). A bibliometric analysis of solid waste research during the period 1993–2008. *Waste Management*, *30*(12), 2410-2417. https://doi.org/10.1016/j.wasman.2010.06.008

González-Alcaide, G. (2021). Bibliometric studies outside the information science and library science field: uncontainable or uncontrollable? *Scientometrics*, in press. https://doi.org/10.1007/s11192-021-04061-3

Harzing, A. W. (2015). Health warning: might contain multiple personalities—the problem of homonyms in Thomson Reuters Essential Science Indicators. *Scientometrics*, *105*(3), 2259-2270. https://doi.org/10.1007/s11192-015-1699-y

Heneberg, P. (2011a). Supposed steep increase in publications on cruciate ligament and other topics. *European Journal of Orthopaedic Surgery & Traumatology*, 21(6), 401-405. https://doi.org/10.1007/s00590-010-0722-5

Heneberg, P. (2011b). On bibliometric analysis of Chinese research on cyclization, MALDI-TOF,

and antibiotics: methodical concerns. *Journal of Chemical Information and Modeling*, *51*(1), 1-2. https://doi.org/10.1021/ci100162w

Heneberg, P. (2013). Lifting the fog of scientometric research artifacts: On the scientometric analysis of environmental tobacco smoke research. *Journal of the American Society for Information Science and Technology*, 64(2), 334-344. https://doi.org/10.1002/asi.22753

Ho, Y. S. (2013). Comments on "a bibliometric study of earthquake research: 1900–2010". *Scientometrics*, 96(3), 929-931. https://doi.org/10.1007/s11192-012-0915-2

Jacsó, P. (2009). Errors of omission and their implications for computing scientometric measures in evaluating the publishing productivity and impact of countries. *Online Information Review*, *33*(2), 376-385. https://doi.org/10.1108/14684520910951276

Li, K., Rollins, J., & Yan, E. (2018). Web of Science use in published research and review papers 1997–2017: A selective, dynamic, cross-domain, content-based analysis. *Scientometrics*, *115*(1), 1-20. https://doi.org/10.1007/s11192-017-2622-5

Liu, W. (2017). The changing role of non-English papers in scholarly communication: Evidence from Web of Science's three journal citation indexes. *Learned Publishing*, 30(2), 115-123. https://doi.org/10.1002/leap.1089

Liu, W. (2019). The data source of this study is Web of Science Core Collection? Not enough. *Scientometrics*, *121*(3), 1815-1824. https://doi.org/10.1007/s11192-019-03238-1

Liu, W., Hu, G., & Tang, L. (2018). Missing author address information in Web of Science—An explorative study. *Journal of Informetrics*, *12*(3), 985-997. https://doi.org/10.1016/j.joi.2018.07.008

Liu, W., Tang, L., & Hu, G. (2020). Funding information in Web of Science: An updated overview. *Scientometrics*, *122*(3), 1509-1524. https://doi.org/10.1007/s11192-020-03362-3

Paul-Hus, A., Desrochers, N., & Costas, R. (2016). Characterization, description, and considerations for the use of funding acknowledgement data in Web of Science. *Scientometrics*, 108(1), 167-182. https://doi.org/10.1007/s11192-016-1953-y

Pautasso, M. (2014). The jump in network ecology research between 1990 and 1991 is a Web of Science artefact. *Ecological Modelling*, *286*, 11-12. https://doi.org/10.1016/j.ecolmodel.2014.04.020

Tang, L., Hu, G., & Liu, W. (2017). Funding acknowledgment analysis: Queries and caveats. *Journal of the Association for Information Science and Technology*, *68*(3), 790-794. https://doi.org/10.1002/asi.23713

Tang, L., & Walsh, J. (2010). Bibliometric fingerprints: name disambiguation based on approximate structure equivalence of cognitive maps. *Scientometrics*, *84*(3), 763-784. https://doi.org/10.1007/s11192-010-0196-6

Vera-Baceta, M. A., Thelwall, M., & Kousha, K. (2019). Web of Science and Scopus language coverage. *Scientometrics*, *121*(3), 1803-1813. https://doi.org/10.1007/s11192-019-03264-z

Zhu, J., & Liu, W. (2020). A tale of two databases: the use of Web of Science and Scopus in academic papers. *Scientometrics*, *123*(1), 321-335. https://doi.org/10.1007/s11192-020-03387-8